\documentclass[%
 aip,
 jmp,%
 amsmath,amssymb,
 reprint,%
]{revtex4-1}

\usepackage{graphicx}
\usepackage{dcolumn}
\usepackage{bm}
\usepackage{upgreek} 
\usepackage{url}

\begin{document}
\title{A Distributed GUI-based Computer Control System for Atomic Physics Experiments}

\author{Aviv Keshet, Wolfgang Ketterle}
\affiliation{Massachusetts Institute of Technology, Department of Physics, Research Laboratory of Electronics, MIT/Harvard Center for Ultracold Atoms. 77 Massachusetts Ave, Cambridge, MA, 02139}

\date{\today}
\begin{abstract}
Atomic physics experiments often require a complex sequence of precisely timed computer controlled events. A distributed GUI-based control system designed with such experiments in mind, The Cicero Word Generator, is described. The system makes use of a client-server separation between a user interface for sequence design and a set of output hardware servers. Output hardware servers are designed to use standard National Instruments output cards, but the client-server nature allows this to be extended to other output hardware. Output sequences running on multiple servers and output cards can be synchronized using a shared clock. By using an FPGA-generated variable frequency clock, redundant buffers can be dramatically shortened, and a time resolution of 100ns achieved over effectively arbitrary sequence lengths.

\end{abstract}

\pacs{07.05.Dz, 07.05.Hd}
\keywords{computer control, control systems, data acquisition}
\maketitle

\section{\label{sec:intro}Introduction}
The study of Bose-Einstein condensates and degenerate Fermi gasses of trapped atoms are one of the most active and exciting sub-fields of atomic physics. Experiments require a sophisticated combination of vacuum, electronic, and laser technology. Most probes of condensates and degenerate gasses are destructive, so data is acquired by repeated ``shots" in which a sample is prepared and then probed. A single shot in such experiments takes $\sim$10-60s to acquire, and requires several hundred precisely timed events, such as opening and closing of laser shutters, ramping and switching magnetic coils, RF evaporation sweeps, and camera triggers. Events may be as long as several seconds (the loading of a magneto-optic trap, for instance), or as short as a microsecond (a blast of laser light to remove unwanted atoms from a trap, for instance). Thus, these experiments require a computer control system capable of outputting a precisely sequenced set of outputs over a large number of analog and digital channels.

Such experiments are also by nature permanent prototypes, in a constant state of being upgraded, tweaked, repaired, and improved. Thus, it is desirable to have a computer control system that is intuitive to use, allowing for easy comprehension, design and modification of output sequence by users who are not experts in the control system's inner workings.

This paper describes a graphical-user-interface-based distributed computer control system developed at MIT for our experiments with ultracold atoms, called the Cicero Word Generator. The software has been used in Fermi gas experiments in the primary author's lab\cite{Sanner:2010hq, Sanner:2011jp, Sanner:2012gf}. In addition, the package (and its source code) are freely available for download and use by other groups\cite{ciceroGithub}, and has been adopted in a number of atomic physics experiments\cite{Campbell:2010je, Yefsah:2011gy, Jo:2009jj, Schirotzek:2009tq, Schunck:2008vk, Heo:2011cf} in groups at over 10 institutions. While designed with BEC and Fermi gas experiments in mind, it is likely that Cicero (or ideas in its design and implementation) could be useful in other types of experiments where elaborate and precise output sequences are required. 

Our system provides outputs both on hardware clocked channels, where precise ($\sim$10ns) timing without shot-to-shot variation is required, as well as on software clocked channels such as GPIB and RS232 (serial) interfaces where deterministic timing is less feasible and generally not required. For deterministic outputs, we use commercial National Instruments (NI) output cards, with the ability to use an FPGA to generate a synchronization signal that allows us to reach time resolutions of $\sim$100ns over effectively arbitrary length sequences. The deterministic output configuration is discussed in Section~\ref{sec:hardtime}. GPIB and RS232 outputs, which run in less reliable software time, are described in Section~\ref{sec:softtime}.

Many experimenters in the field end up writing their own control software in-house, often in isolation from other groups, leading to a large duplication of effort. The authors are aware of a few other published accounts of control software systems that have been shared between institutions \cite{shreck, owenHall, steck, lewandowski}. This work is complementary to those, and is distinguished by the fact that it takes a graphical user interface approach to designing sequences (rather than the text-based sequence programming approach offered by others) and by its targeting of commercially available NI output hardware rather than custom made parts. There is a tradeoff between a potentially greater versatility and automation in the programming-interface approach, versus greater ease of use and comprehensibility of the graphical-interface approach, though we attempt to address this with certain advanced features of the graphical approach to be explained in \ref{sec:ui:cicero}.

An overview of the control system is presented in Section~\ref{sec:arch}. The user interface is described in Section~\ref{sec:ui}. The details of timing and synchronization schemes are described in Section~\ref{sec:outputs}.

\section{\label{sec:arch}Architecture}

\begin{figure*}
\includegraphics{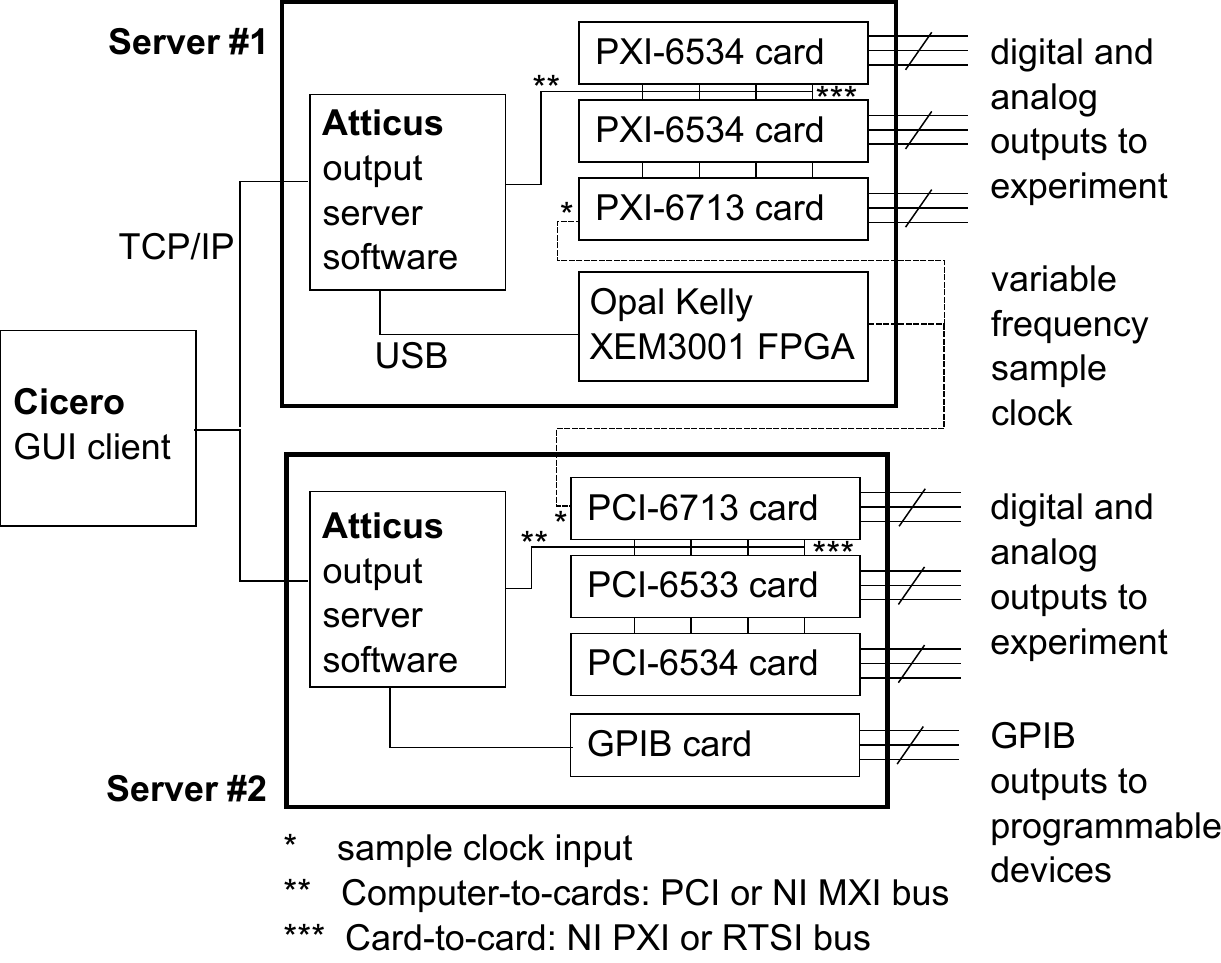}
\caption{\label{fig:architecture}A typical installation of Cicero and Atticus, with two output servers. In this case, an FPGA is being used to synthesize a sample clock used to synchronize output channels.}
\end{figure*}

Cicero splits the job of designing and running output sequences using a client-server architecture. A typical Cicero and Atticus installation is depicted in Figure~\ref{fig:architecture}. The client, generally referred to as Cicero\footnote{both the full software suite and the user interface client are generally referred to by the same name, Cicero, but the intended meaning is usually clear from context}, provides a graphical user interface for loading, saving, and editing sequences, and for starting runs. The server, Atticus, handles the output hardware configuration and converts high-level Cicero sequence objects into an output buffer for the output hardware located on the server computer. The client communicates with one or more servers over a standard TCP/IP network. 

Splitting the output hardware from the user interface gives several advantages: it allows the system to scale to large numbers of output channels -- more than could be supported by a single computer; it allows the system to be generalized to run on other types of hardware, without modification of the graphical user interface, by making other server implementations for different output hardware; and it allows for output hardware to be physically located close to its point of use, rather than necessarily being close to the experiment operator's computer.

An individual shot breaks down into several steps. First, Cicero sends to each server a high level sequence description. The servers turn the sequence description into output buffers for each of the channels hosted by that server. These buffers are loaded into the output card memory, and the cards are armed to begin output. Cicero then sends a trigger command to the servers, which depending on the configuration either initiates a shared sample clock, or outputs a trigger signal to cards using their own internal clocks. After the sequence has run, Cicero polls the servers for any errors encountered in the run. If no error was encountered, and Cicero is set to loop or scan over a parameter list, then Cicero repeats the process for the next shot.

\section{\label{sec:ui}User Interface}
\subsection{\label{sec:ui:cicero}Client -- Cicero}
Cicero is a descendant of commercial Word Generator control hardware used in early atom cooling and trapping experiments at MIT, which through front-panel programming allowed users to pre-program and then run sequences of synchronized digital outputs over a collection of channels. These manually programmed devices were eventually succeeded by several generations of computerized control systems, based on custom software and computer integrated output cards, but retaining the basic sequence-of-words scheme and the historical Word Generator name.

\begin{figure*}
\includegraphics{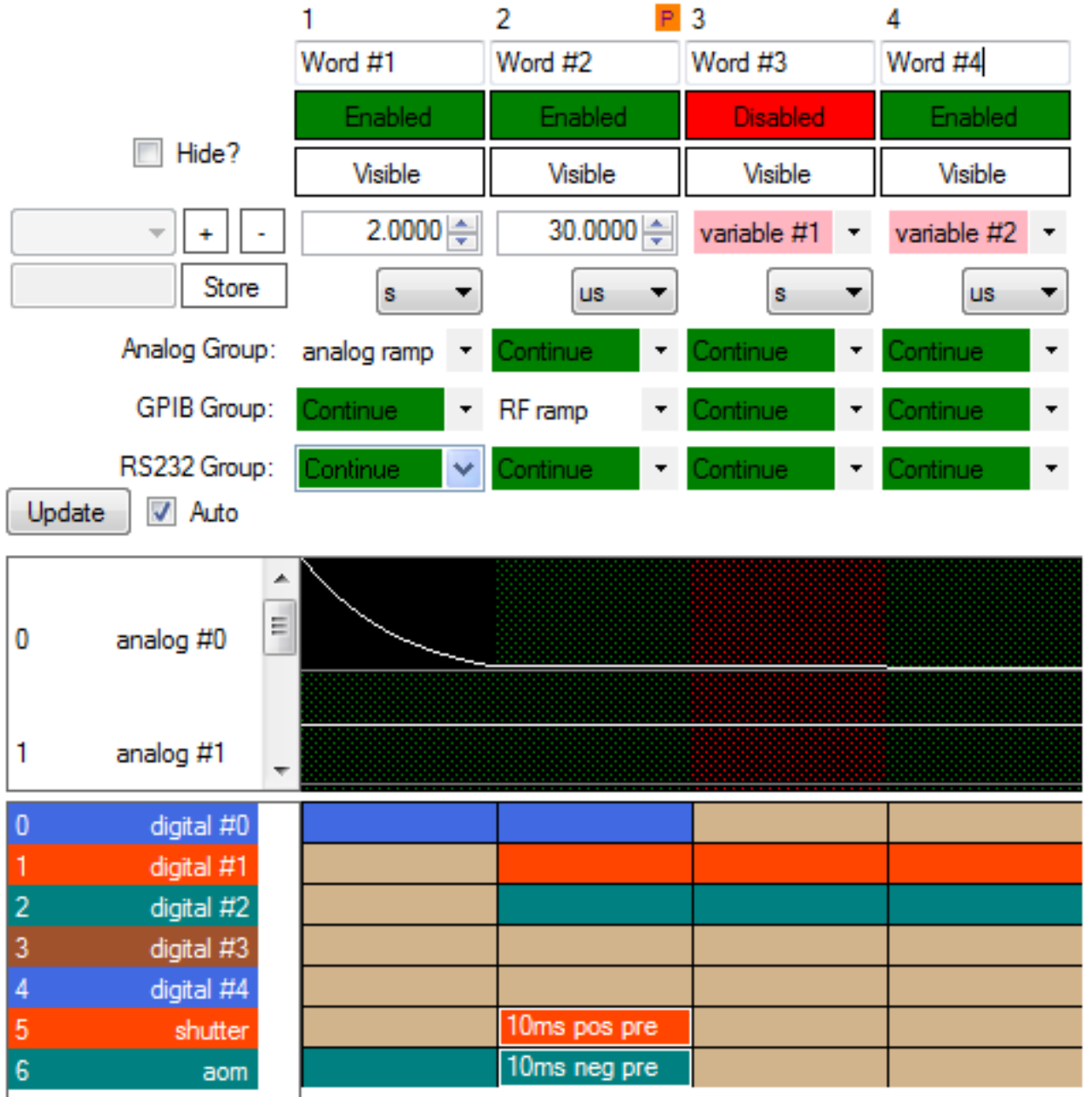}
\caption{\label{fig:loremIpsum}Screenshot of the main sequence editing user interface.}
\end{figure*}

A screenshot of Cicero's basic sequence editing user interface is depicted in Figure~\ref{fig:loremIpsum}. A sequence consists of a series of words -- columns of user-settable time duration. In each word, the user specifies the output value of digital channels using a grid of toggles. A word may optionally trigger ramps of analog output channels, or trigger GPIB or RS232 output commands. 

The essential elements of the user interface were carried over from previous generations of control software. The interface is intuitive to people with no programming experience, and by glancing at the sequence editing screen it is easy to see which channels are doing what when. Two key user interface enhancements have been introduced to alleviate many of the disadvantages of the GUI approach to the programming interface approach. 

In our previous implementations of this basic GUI approach, it was extremely tedious to repeat a sequence while systematically changing a parameter, a task very common when collecting data or optimizing the apparatus. This has been alleviated with the introduction of variable parameters. Any numerical parameter in a sequence can be bound to a variable by right-clicking on it and selecting from a menu of defined variables. Variables can be assigned hard values, or defined in terms of other variables by entering mathematical formulae, or assigned to lists which can be scanned over, allowing the variable to take on a succession of values in successive iterations of the sequence.

In our experience with these experiments and previous GUI implementations, actions which are logically a single operation often required several words in the GUI to accomplish. A classic example is flashing on an imaging laser beam for a short pulse. Such beams are typically controlled by a combination of an acousto-optic modulator (very fast rise time, imperfect extinction ratio) and mechanical shutter (slow rise, perfect extinction ratio). The modulator must be kept pre-warmed for about a second before the pulse, but must be kept off during the slow shutter rise time ($\sim$10ms).  This meant every imaging pulse word was accompanied by a pre-trigger word which turned off the modulator but began opening the shutter. Such pre-triggers could be extremely complicated if they overlapped with other pre-trigger words or other time sensitive phases of a sequence (such as the release of atoms from trap for a time-of-flight image). These problems have been alleviated in Cicero by introducing the ability to define more sophisticated digital actions than just turning on during a word. These actions are termed Pulses, and are created and edited in a special section of the GUI. They support a variety a pre- or post-trigger behavior, in essence allowing them to cause digital channel value changes at times that are specified relative to the boundaries of the word that the Pulse is placed in, but not confined to the word boundaries. Pulses are then assigned to an individual channel at given word by right clicking on the grid of toggles (see word 2 of Figure~\ref{fig:loremIpsum}). Using these pre-triggers can help ensure that each word of the sequence corresponds to one logical operation, making the sequence more modular, easier to read, and making it much less tedious to accomplish sequences with overlapping pre-triggers.

A slew of other client features have been incorporated as they have become needed, and an exhaustive description of them is beyond this paper's scope, but they include: the ability to loop a keep-warm sequence in the background while editing a sequence in the foreground; grouping a set of words into a module, allowing them to be batch-enabled, -disabled or looped; ramps and waveforms defined graphically or symbolically; logging of run details to explorable log files or to a database (SQL); persistent variables that can be referenced from multiple sequence files; and interspersing of calibration shots into a long set of parameter scanning runs.

\subsection{\label{sec:ui:atticus}Server -- Atticus}

The Atticus output server is quite versatile, able to communicate with a wide range of output hardware, in many possible timing configurations. This requires a certain amount of installation-specific configuration. Unlike much experiment-specific control software, Atticus makes all such configuration settings accessible to the user through a GUI, without needing to make changes to hard-coded parameters in the source code.

\section{\label{sec:outputs}Output Details}
\subsection{\label{sec:hardtime}Synchronization scheme for analog and digital channels}
Our control system is built around NI output cards, specifically the PXI- and PCI- 6713 (12-bit analog outputs, 8 channels) and PXI- and PCI- 6534 (32-bit output digital output). These cards allow for deterministic sequence output, without relying on unreliable timing supplied by a computer operating system.

Generically, all such output cards function in a similar manner. Before running the sequence, an output buffer for each channel is precomputed, containing the value that the channel will take on at each sample time. Part of the buffer resides in the card's on-board memory, while the rest resides on the controlling computer and is streamed to the card as necessary. To run the sequence, a start trigger signal is supplied to the cards, which initiates output generation. A sample clock signal advances each channel to the next sample in its buffer. By physically sharing the start trigger and/or sample clock signals (using a card-to-card bus if the cards are co-located, or a coaxial cable if they are physically  separated) all the output samples of all cards in the experiment can be precisely synchronized.

In the simplest realization, a few cards on a single NI bus can share a sample clock that is generated by one of the cards' on-board oscillators. The time resolution of the output sequence (i.e. the shortest word that the sequence may contain) is then set by the the sample clock frequency. In our experience, some legacy PCI cards (still in use in labs at MIT) can only reliable sustain continuous sample generation at rates up to $\sim$50kHz. At rates above this, streaming from the computer memory to the card is not always able to fill the on-board buffer's fast enough. This translates to a shortest word size of 20$\upmu$s.

In a typical sequence, a minority of the words in which no channel values are changing take up the majority of the sequence time (for instance during MOT loading or RF evaporation sweeps), but it is still desirable to have a few very short words for fast operations. When used with the naive fixed-frequency sample clock described above, the situation is the worst of two worlds. The shortest word size is inconveniently long compared to some of the fastest operations we may want to perform, but the vast majority of the buffer is filled with redundant repeated samples during long words just to achieve a high time resolution, meaning the buffer is both large and slow to generate, as well as insufficiently high resolution. The solution to these problems is to use a sample clock that is not fixed frequency.

Instead of a fixed frequency clock, a variable frequency clock can be synthesized, one that has edges only when the sequence calls for changes in the output values. This synthesized clock can then be sent to the sample clock input of the output cards, and allows the buffers on those cards to be small even while the time resolution of the sequence is high. We have developed two schemes for creating a variable frequency clock, using either a normal digital output from an NI card, or using an FPGA. In both schemes, for simplicity, all output cards share the same sample clock, so that if any channel on any card needs to change values, all cards are updated. In addition, a clock edge will occur at each word boundary even if no outputs change. In practice, this reduces complications in the code while causing only a limited amount of redundant buffer generation.

NI Cards (such as PXI-6534) typically have two halves which can make use of separate output buffers and sample clocks. Half of a card can be sacrificed to provide a variable frequency clock output. If set up for this form of clock generation, a single large buffer for this synthetic clock is calculated for each sequence run, and the appropriate digital output of the card is then fed into the sample clock inputs of other cards. The synthetic clock buffer itself  is clocked using the on-board oscillator of the card, typically at 1MHz, achieving an effective time resolution in the output sequence of 2$\upmu$s (since the shortest interval in the output sequence requires at least two samples of the clock sequence, one positive trigger edge and one falling edge). The scheme works, but has downsides: Half of a digital output card has to be sacrificed; the very large synthetic clock buffer can take several seconds to generate for each shot, and requires a card with a large on-board buffer.

An alternate approach is to use an FPGA to synthesize the variable frequency clock. We have made use of the XEM3001 FPGA board available from Opal Kelly, which is inexpensive, simple, and provides an easy to use computer-USB-FPGA communication interface. The FPGA is programmed on each Atticus startup with custom clock-synthesis code. Instead of creating the large output buffer necessary for the output-card-synthesized clock, when using an FPGA Atticus merely computes a list of clock frequencies and dwell times for these frequencies. This list is transferred before each run to the FPGA over a USB connection. When the run is triggered, the FPGA begins synthesizing a variable frequency clock on the fly, by counting down from either its on-board 10MHz oscillator or from an externally provided one. This translates to a time resolution of 100ns, (with a shortest word length of 200ns), vastly finer than needed in our experiments. None of the shortcomings of the synthetic clock buffer apply, since no large buffer needs to be precomputed.

Using an FPGA as the synchronization source also enables a basic form of real-time sequence feedback, namely the ability to pause and retrigger a run in response to some measurement. In the Cicero GUI, a given word can be marked with a ``Hold then Retrigger" flag. After the FPGA reaches the part of the clock generation that corresponds to the beginning of this word, it pauses until receiving a retrigger signal on a dedicated digital input. Doing this retriggering in hardware with an FPGA (rather than in software, by having Atticus attempt to pause and resume output cards) allows it to be as fast and precise as any other hardware-timed event. Sequence retriggering of this type can help reduce shot-to-shot fluctuations caused by the environment, for instance by triggering certain parts of the experiment to coincide with a given phase of the AC mains line, or with the number of atoms loaded into a MOT as measured by fluorescence detection.

\subsection{\label{sec:softtime}GPIB, Serial, and other output}

In addition to the various digital and analog signals synthesized by output cards, laser cooling experiments often make use of programmable function generators and synthesizers to produce RF and microwave sweeps. A common and obvious example is RF evaporation, in which a synthesizer frequency must be precisely swept over a range of frequencies, often with a precisely tuned and non-uniform
ramp profile. With legacy synthesizers, such as those from the Agilent ESG series, this is accomplished by issuing to the device a time-series of GPIB commands jumping the device to the desired frequency. Newer synthesizers, such as the NI-RFSG, accept commands from the computer using a much faster PXI interface. Many other function generators and translation stages accept commands over a RS232 (``Serial") port.

What all these communication methods have in common is that they require a computer to send the command at the correct time in the sequence, at least to the best ability of a computer clock within the limitations of a non real-time operating system (though the authors recently became aware of a triggerable GPIB controller which can be pre-programmed in much the same way as our analog and digital outputs\cite{hallGpib}).

These outputs are described as software clocked, to distinguish them from
the hardware clocked outputs which do not rely on the computer's concept of time.
Simply relying on a computer's on-board clock is often acceptable. The timing jitter of such an approach is generally less than 10ms, which is good enough considering that the GPIB command latency of a typical Agilent ESG synthesizer is over 100ms. In this mode, Atticus starts a thread at the beginning of a sequence run which continuously polls the computer's on-board clock. This is used to trigger the output of the correct GPIB, serial, or RFSG commands at the correct time in the sequence run.

In some circumstances, this timing scheme fails. When an FPGA clock is being used, along with the above described retriggering feature, any time the FPGA spends waiting for a retrigger translates directly into a skew between the software and hardware clocked outputs. If the retrigger waits are long, this can be an unacceptable long skew. Thus, Atticus can be conguring to use a different method of software clocking -- FPGA polling. In this mode, Atticus continuously polls the FPGA for its accurate sequence time (which takes
into account any pauses and retrigger waits). This FPGA polled time is then used instead of the computer's on board clock to determine when commands are output. When configured to do so, the FPGA-derived software clock can be broadcast over the network using a lightweight UDP stream, both to the Cicero client so that the user interface can have a more accurate display of the sequence position, and to other Atticus hardware servers so that the software clocked events on that server can also be kept synchronized to the hardware clocked ones.

\section{Conclusion}

The Cicero Word Generator control system provides a user-friendly and powerful solution to the problem of running elaborate output sequences for atomic physics experiments. This system should be generalizable to other types of experiments where realtime sequence feedback (beyond pausing and resuming an output sequence) is not required, and where a rapid and intuitive sequence editing user interface is desired. The separation of the software into a user interface program and an output program should also allow the system to be used with output hardware other that that described here, by writing a custom server implementation. It is the authors' hope that the software itself, or ideas described here, can be of use to other experimentalists.

\begin{acknowledgments}
The authors wish to acknowledge Widagdo Setiawan for planting the seed for this project, and Tony Kim, Ryan Olf, Emmanuel Mimoun, Edward Su, and Timur Rvachov for contributions and useful discussions. This work was supported by NSF and ONR, AFOSR MURI, and under ARO Grant No. W911NF-07-1-0493 with funds from the DARPA Optical Lattice Emulator program.
\end{acknowledgments}

\bibliography{cicero}

\end{document}